# Tuning the electrical conductivity of Pt-containing granular metals by postgrowth electron irradiation

F. Porrati1, R. Sachser1, C. H. Schwalb1, A. S. Frangakis2 and M. Huth1

1. Physikalisches Institut, Goethe-Universitäat, Max-von-Laue-Str. 1, D-60438 Frankfurt am Main, Germany
2. Institut für Biophysik, Goethe-Universität, Max-von-Laue-Str. 1, D-60438 Frankfurt am Main, Germany

**Abstract**

We have fabricated Pt-containing granular metals by focused electron beam induced deposition from the $(CH_3)_3CH_3C_5H_4Pt$ precursor gas. The granular metals are made of platinum nanocrystallites embedded in a carbonaceous matrix. We have exposed the asgrown nanocomposites to low energy electron beam irradiation and we have measured the electrical conductivity as a function of the irradiation dose. Postgrowth electron beam irradiation transforms the matrix microstructure and thus the strength of the tunneling coupling between Pt nanocrystallites. For as-grown samples (weak tunnel coupling regime) we find that the temperature dependence of the electrical conductivity follows the stretched exponential behavior characteristic of the correlated variable-range hopping transport regime. For briefly irradiated samples (strong tunnel coupling regime) the electrical conductivity is tuned across the metal-insulator transition. For long-time irradiated samples the electrical conductivity behaves like that of a metal. In order to further analyze changes of the microstructure as a function of the electron irradiation dose we have carried out transmission electron microscope (TEM), micro-Raman and atomic force microscopy (AFM) investigations. TEM pictures reveal that the crystallites' size of long-time irradiated samples is larger than that of as-grown samples. Furthermore we do not have evidence of microstructural changes in briefly irradiated samples. By means of micro-Raman we find that by increasing the irradiation dose the matrix changes following a graphitization trajectory between amorphous carbon and nanocrystalline graphite. Finally, by means of AFM measurements we observe a reduction of the volume of the samples with increasing irradiation time which we attribute to the removal of carbon molecules.

## 1. Introduction

Granular metals are nanocomposite materials made of metallic grains of few nanometer size embedded in a dielectric matrix. These materials experience a renewed interest which is drawn by the technical advances in manipulating the matter at nanometer scale [1, 2]. In parallel, by tuning the size and the location of the metallic grains and by controlling the characteristic of the dielectric matrix granular metals can be used to study the interplay of electronic correlations, quantum size effects and disorder [3]. The electronic transport properties of granular metals are governed by the (inelastic) tunneling between metallic grains through the potential associated with the dielectric matrix. In accordance with the strength of the intergrain tunneling granular metals show either non-insulating or insulating behavior [4, 5]. For mid intergrain coupling, due to the ability of fine tuning, granular metals are considered to be ideal model systems to study the metal-insulator transition (MIT) [3, 6, 7].

In the last years, among the techniques used to prepare granular metals focused electron beam induced deposition (FEBID) has received much attention [2, 8]. FEBID is a versatile direct-write technique employed to fabricate samples at nanoscale. By interaction with the electron beam of a scanning electron microscope (SEM) the adsorbed molecules of a precursor gas injected in the SEM dissociate into a volatile and a nonvolatile part, the deposit, which constitutes the sample. The most used precursors are metal carbonyls and organometallic compounds. Therefore, a large amount of carbon is present in the deposits which are made of metallic nanoparticles embedded in a carbonaceous matrix. Trying to increase the metal content at the expense of carbon is required to obtain high conductivity samples for nanotechnology applications [9].

The aim of the present work is to investigate the effect of the postgrowth electron irradiation on deposits fabricated by FEBID. As precursor gas we choose the trimethyl (methylcyclopentadienyl)platinum, $(CH_3)_3CH_3C_5H_4Pt$. The deposits grown from this precursor gas contain Pt nanocrystallites embedded in an amorphous carbon matrix [10, 11]. In most instances, the effect of electron irradiation on carbon nanostructures is considered to be detrimental because the incident electrons can remove carbon atoms damaging the sample [12]. However, irradiation can also have beneficial effects. For instance, the morphology of carbon nanotubes, graphene and fullerene can be controlled by electron beam irradiation [13, 14]. Moreover, electron irradiation has been successfully used to tailor mechanical and electronic properties of carbon nanomaterials [15, 16]. In the present work we report on the effect of low energy electron beam irradiation on the Pt-containing deposits. In particular, we show that the electronic structure of the carbonbased matrix can be tuned by the irradiation dose. As a consequence, the tunneling coupling between Pt nanocrystallites changes and the electron transport regime can be tuned from insulating to metallic.

## 2. Experimental details

To prepare the samples we have used a dual-beam SEM/FIB microscope (FEI, Nova Nanolab 600) equipped with a Schottky electron emitter. The precursor gas was introduced by a gas injection system via a 0.5 mm diameter capillary in close proximity to the focus of the electron beam on the substrate surface. The samples were grown on a Si(p-doped)/SiO2 (100 nm) substrate. The contact electrodes were prepared by UV photolithography with 120 nm Au/Cr, see Fig. 1a. The sample were obtained by repeatedly rastering the area of interest. The rastering constant, or pitch, was chosen to be 20 nm in order to obtain a homogeneous deposit [17]. The typical area of the deposits is 15 m2, see Fig. 1b. The thickness is around 80 nm in the as-grown state, as revealed by AFM measurements (Fig. 1c). The deposits were prepared with a beam energy, beam current and dwell time of 5 keV, 1.6 nA and 1 s, respectively. The typical composition of the deposits is ca. 24 at% Pt and 76 at% C, as measured by energy dispersive x-ray spectroscopy (EDX). The postgrowth irradiation experiments were performed with the same beam parameters, with variable exposure time. For micro-Raman measurements we prepared samples with area 4x4 μ2 and thickness around 160 nm. For TEM investigations we grew samples with area of 15 m2 on electron transparent carbon membranes with a thickness of about 10 nm.

For the electrical conductivity measurements in the temperature range 1.8-265 K we used a variable-temperature insert mounted in a 4He cryostat . Standard measurements were performed with a Keithley Sourcemeter 2400 at fixed bias voltage, which resulted in an electrical field of about 30 V cm−1. For low current measurements a Keithley Sourcemeter 2636A was employed. For AFM measurements in non-contact mode we used a commercial microscope (Nanosurf, easyscan2) equipped with a tip with radius around 7 nm, height 10-15 μm and half cone angle at the tip apex of around 10◦ Micro-Raman measurements were carried out with a Ranishaw spectrometer with a He-Ne laser of 633 nm wavelength and 50 mW max. output power. The measurements were made with 5 mW power to avoid undesired sample annealing or even damage. TEM investigations were performed with a FEI Tecnai F20 microscope at 200 keV.

## 3. Results

### 3.1 Electrical conductivity measurements

In Fig. 2b we report a representative current vs. time measurement at fixed sample bias voltage obtained during electron irradiation. The curve refers to a sample irradiated for 7 hours, which is the longest irradiation time used in the investigation. As mentioned above, sample preparation and irradiation were carried out with beam energy and beam current of 5 keV and 1.6 nA, respectively. The current increases with time, showing that the intergrain tunneling probability increases by irradiation dose. Furthermore, the current slope is not constant during the irradiation process, which indicates a complex microstructural transformation. In section 4 we discuss possible reasons for this behavior.

In Fig. 3 we report temperature-dependent conductivity measurements performed in the range 1.8 K to 260 K. We show four curves, each of them signature of a different electronic transport regime. The as-grown sample behaves like an insulator. In the range 19 K to 260 K the temperature-dependent conductivity follows a stretched exponential of the form $s \sim \exp(-T_0/T)^{(1/2)}$, which is characteristic of the correlated variable-range hopping (VRH) transport regime in three dimensions [18]. Recently, this behavior has been reported for FEBID deposits grown from $W(CO)_6$ and $(CH_3)_3CH_3C_5H_4Pt$ precursor gases [19–21]. In particular Pt-containing deposits showed VRH behavior in the measured range 70-300 K [20]. The temperature-dependent conductivity of the sample irradiated for 500 seconds exhibits a saddle point at about 30 K. This behavior is characteristic of samples with mid intergranular coupling close to the MIT, as reported for homogeneous samples and 2D nanodots arrays prepared by FEBID from $W(CO)_6$ precursor gas [19, 22, 23]. For the sample irradiated 80 minutes, the electronic transport is governed by strong intergranular coupling. The conductivity tends to a finite value in the limit of T to 0, as expected for samples on the metallic side of the MIT [3]. Finally, the sample irradiated for 7 hours shows a behavior conventionally associated to metallic conduction, namely a decreasing conductivity with growing temperature.

One effect of the electron irradiation on the as-grown sample is to reduce the thickness of the sample, as is evident from Fig. 1c and 1d. The depression shown is caused by a few minutes of local electron irradiation during an EDX measurement. In order to study systematically the dependence of the thickness on the irradiation dose, we have carried out AFM measurements of the series of samples irradiated in the time range from 100 s to 7 h, see Fig. 2a. In Fig. 4a we report the measured thickness vs. the irradiation time. This dependency can be used to calculate the electrical conductivity during irradiation. By dividing the curve of Fig. 2b with the one of Fig 4a we obtain the electrical conductivity, see Fig. 4b. With the dots we plot the conductivity of the individual samples, obtained by measuring the current-voltage characteristic for each sample after irradiation. The points nicely follows the curve, showing that the irradiation-dependent conductivity is universal for each sample. As a consequence, we can assume that the same microstructural changes take place with the same irradiation dose for the different samples.

**3.2 Micro-Raman and TEM measurements**

In Fig. 5 we plot the micro-Raman spectra obtained for an as-grown sample (a) and two samples irradiated with $1.42 \times 10^5$ C/m2 dose (b) and $2.84 \times 10^5$ C/m2 dose (c), respectively. We find two main peaks at around 1375 cm−1, the so called "D" peak, and at around 1550 cm−1, the so called "G" peak. The D and G peaks are associated, respectively, with the breathing and stretching modes of the sp2 carbon bonds [24]. In the case of our deposits the D and G peaks refer to the carbonaceous matrix. Note that the total number of counts decreases with the irradiation time. This is due to the matrix volume reduction under irradiation in comparison with the Pt particles' total volume. Recently, a similar decrease has been reported for metal-doped diamond-like carbon thin films [25]. The analysis of the position and of the intensity of the peaks gives indications for the microstructural configuration of the matrix and the dominant carbon bond type, i.e. sp2 and sp3 [24]. In Fig. 6 we report the position, the ratio of the intensities and of the areas of the peaks D and G. All these parameters increase with the irradiation time. In particular, the position of the G peak lies between 1542 cm−1 and 1563 cm−1 and the ratio of the intensities is 0.89 <= I(D)/I(G) <= 1.81. Therefore, according to Ref. [24], by increasing irradiation time the matrix of the deposits follows a graphitization trajectory between amorphous carbon and nanocrystalline graphite. In parallel, the number of sp3 (sp2) bonds decreases (increases) from around 10% (90%) to around 5% (95%).
The TEM investigations were carried out for one non-irradiated sample and for five post-irradiated samples. In order to visualize directly the effect of the irradiation, we have irradiated only a part of the sample's area. The irradiation doses used correspond to 250 s, 500 s, 30 min, 80 min and 7 h irradiation time (see Fig. 2). On the left side of Fig. 7 we show a TEM image of the transition between irradiated and non-irradiated areas. The image refers to the longest irradiated sample. One can notice a difference between the granularity in the irradiated region (a1) and the one in the non-irradiated region (b1). This difference is more evident by comparing the selected zoom of the irradiated region (a2) with the one of the non-irradiated region (b2). In particular, the mean crystallite size in the irradiated region appears to be larger than in the non-irradiated region. Furthermore, some crystallites coalescence takes place as shown in the regions bordered in yellow. On the right side of Fig. 7 we present the diffraction pattern corresponding to the two regions. These images show that the sharpness of the diffraction rings increases with irradiation, which can be due to both an increase of the crystallinity and of the grain size. The same analysis reported in Fig 7 for the longest irradiated sample has been carried out also for the other samples. In the TEM images of these samples we do not find any difference between non-irradiated and irradiated areas, meaning that in these cases the grain size is not affected from the irradiation.

## 4. Discussion

The ability to fine control the electrical conductivity of granular metals grown by FEBID is important for fundamental studies and applications alike. In the literature various methods are discussed to obtain an increase of the electrical conductivity [9]. Among these methods are: careful selection of the electron beam parameters used for the deposition [26], annealing of the samples in reactive atmosphere [27] and deposition from carbon-free precursors [28]. Postgrowth electron beam irradiation has been used to tailor the grain size of Pt nanocrystallites and to increase the electrical conductivity in deposits grown from the carbon-free $Pt(PF_3)_4$ precursor [29]. In other studies it was shown that postgrowth electron irradiation carried out in a transmission electron microscope (TEM) promotes the increase of the crystallinity in deposits prepared from $Pt(PF_3)_4$ and $W(CO)_6$ precursors [31, 32]. In the present work we show that the electrical conductivity of Pt-containing deposits from $(CH_3)_3CH_3C_5H_4Pt$ precursor can be tuned by low energy electron irradiation. The electrical conductivity increases by 3 orders of magnitude during the first 30 minutes of electron irradiation, see Fig. 4. This is a very significant value if compared with the 30 times increase of Pt-deposits from $Pt(PF_3)_4$ precursor [29]. Moreover, $Pt(PF_3)_4$-derived deposits show saturation effects in the conductivity after 30 minutes of irradiation [29], while in our experiments the conductivity grows further (Fig. 4). The highest value of the electrical conductivity measured after irradiation is $4.6 \times 10^5$ $\Omega^{-1}m^{-1}$ for Pt-deposits from $Pt(PF_3)_4$ precursor [29]. Note that the bulk value for Pt is $0.9 \times 10^7$ $\Omega^{-1}m^{-1}$. In our irradiation experiment we obtained $1.28 \times 10^5$ $\Omega^{-1}m^{-1}$ for the sample with maximum irradiation time. This value is even larger than the one obtained after the annealing procedure in reactive atmosphere ($0.7 \times 10^4$ $\Omega^{-1}m^{-1}$) proposed in Ref. [27] and the largest, to our knowledge, reported for Pt-containing deposits from the $(CH_3)_3CH_3C_5H_4Pt$ precursor. In our experiment the total increase of the electrical conductivity amounts to 4 orders of magnitude.

Recently, the MIT has been studied in W-based granular metals fabricated by FEBID from $W(CO)_6$ precursor gas [19]. The metal content of the deposits prepared from this precursor can be strongly influenced by the electron beam parameters. This large tunability allows to reach the metal content necessary to study the MIT. In Ref. [19] the metal content was varied between 19 at% and 36.9 at% by tuning the deposition electron beam parameters. The MIT takes place at about 31 at%. This route is not practicable to reach the MIT in Pt-based deposits fabricated from $(CH_3)_3CH_3C_5H_4Pt$ precursor because in the as-grown state the deposits with highest metal content (24 at%) show insulating behavior. Therefore, the postgrowth electron beam irradiation constitutes a valid alternative to study the MIT in this system. In particular, in our experiments we find that the conductivity of the deposits is very reproducible and continuously tunable. The dependence of the electrical conductivity on the postgrowth irradiation is due to a complex microstructural transformation which may, in principle, involve either the Pt nanocrystallites or the carbonaceous matrix or both. In particular, an increase of the tunnel coupling is expected with an increase of the crystallite size, a variation of the dielectric constant of the matrix or a reduction of the distance between Pt nanocrystallite due to carbon removal. Pt-based deposits from the $Pt(PF_3)_4$ precursor investigated by TEM show an increase of the crystallite size during a post-growth electron irradiation treatment [29]. In Pt-based deposits fabricated from $(CH_3)_3CH_3C_5H_4Pt$ the degree of crystallinity, whose increase in some EBID deposits is associated to a grain size increase during growth [30], increases with the electron flux during the growth process [33]. However, unambiguous changes in crystallinity were not seen in TEM data of postgrowth irradiated samples [33]. Therefore, according to the literature the question of the influence of the electron post-irradiation on Pt-based deposits from the $(CH_3)_3CH_3C_5H_4Pt$ precursor remains open. From our data, see Fig. 7, we see that the granularity of long irradiated samples is different from that of as-grown samples. In particular, the crystallite size has increased and coalescence has started. Therefore, the increase of the electrical conductivity can be attributed to these microstructural changes. On the other hand, for briefly and middle irradiated samples from the TEM data we do not have evidence of microstructural changes. Therefore, even if we cannot exclude that higher resolution imaging could show microstructural changes, from our data we are brought to believe that the change in conductivity is due to a change of the matrix. Note also that if only the crystallinity improves (without grain size increase) the conductivity is not expected to change because in granular materials the electrical transport is dominated by the tunneling between grains [3], i.e. the intragrain conductance is much higher than the intergrain conductance.

As discussed in Section 3.3, the micro-Raman investigation shows that the matrix changes because of the electron irradiation. In particular we have found that the matrix tends to graphitize. Accordingly, the number of sp2 carbon bonds increases at the expense of sp3 bonds [24]. A similar increase is reported also for ta-C carbon films by irradiation of 200 keV electron beam [34]. In general, the conductivity of samples with dominant sp2 bonds (graphite) is higher than samples with dominant sp3 bonds (diamond, ta-C carbon). However, in our experiment we attribute the increase of the electrical conductivity to an effective reduction of the tunneling barrier height caused by the transformation of sp3 into sp2 bonds, rather than to an additional transport channel through the graphitic component in the matrix. In fact, if an additional transport channel would be involved as a consequence of the irradiation, we would expect to see a strong discontinuity in the conductivity temperature dependence or in current-voltage characteristics as a function of the irradiation time. Instead we measure a smooth and continuous change in the conductivity (see Fig. 3).

From the micro-Raman investigation, see Section 3.3, we deduce that the number of sp3 carbon bonds decreases, from around 10% to around 5% . This graphitization process involves a reduction in the density of the matrix which we estimate to be between 2 and 5 percent, assuming a linear relation between density and sp3 bonds [35]. The reduction of the density is reflected in a few percent increase of the volume of the deposit. This small effect is hidden by the much stronger volume reduction due to the carbon removal. In the literature, carbon removal was analyzed in freestanding Pt nanowires [36], which were post-irradiated with 200 keV inside a TEM. The origin of carbon removal was attributed to local heating [27, 37] and knock-on [38] effects. In our experiment, these effects cannot be at the origin of carbon removal because the current and energy of the electron beam are too small [8]. Rather, we speculate that the loss of carbon is due to an electron-beam induced reaction of carbon with the residual gas molecules (in particular H2O and O2) present in the SEM chamber at the base pressure of $4\ 10^{-6}$ mbar.

From our investigation, the removal of carbon takes place since the first minutes of electron irradiation and it contributes to the increase of the conductivity, together with the graphitization process. In particular, due to the carbon removal the distance between

Pt nanocrystallites shrinks (volume decreases) and the tunneling probability increases. This process continues during the whole irradiation experiment and is described by the long decay constant of the fit in Fig. 4a. At some point percolation between neighboring nanocrystallites can take place, further increasing the conductivity. We speculate that percolation starts after around 250 minutes of irradiation, where the irradiation time dependent conductivity becomes linear, see Fig. 4b. In summary, we have three processes which contribute to the microstructural transformation of the matrix: graphitization, which takes place in the first part of the irradiation experiment; carbon removal, along the whole irradiation time; and percolation, confined to the second part of the experiment. A fourth process, confined to the first minutes of the experiment, contributes to the rapid initial decrease of the thickness of the deposits, see Fig. 4a. This process cannot be analyzed with the present resolution of the experiment and it is left to future investigations.

## 5. Conclusions

We have prepared Pt-containing granular metals by FEBID from the $(CH_3)_3CH_3C_5H_4Pt$ precursor. The as-grown samples have been electron irradiated on a time scale ranging between 100 seconds and 7 hours. The electrical transport measurements of the as-grown samples and of the briefly irradiated samples reveal a variation of the tunneling coupling between the Pt nanocrystallites, which is attributed to the graphitization of the matrix, as revealed by micro-Raman investigations. TEM pictures show that the crystallite size of long-irradiated samples is larger than that of as-grown samples, while no changes of the microstructure are found in briefly irradiated samples. Both these effects: the variation of the tunneling coupling, for briefly irradiated samples, and the increase of the crystallites size, for long irradiated samples, explain the increase of the electrical conductivity measured in our transport measurement. Within the time of the electron irradiation experiment the value of the electrical conductivity increased 4 orders of magnitude, which allows us to tune the granular metal from the insulating to the metallic transport regimes. The Pt-containing granular metals fabricated and analyzed in this work are attractive both for basic studies and applications. In fact, on the one hand, the ability of fine tuning makes these samples perfect model systems to study, for example, the MIT. On the other hand, the possibility to vary the electrical conductivity from the insulating to the metallic regime opens a large spectra of possible nanotechnology applications. For example, high conductivity samples may be used as contact electrodes, mid conductive samples as tunable strain sensors [39, 40].

## Acknowledgments

The authors aknowledge financial support by the NanoNetzwerkHessen (NNH), the Bundesministerium für Bildung und Forschung (BMBF) under grant no. 0312031C and the Beilstein-Institut, Frankfurt/Main, Germany, within the research collaboration NanoBiC.

## References

[1] C. B. Murray, C. R. Kagan, and M. G. Bawendi, Annu. Rev. Mater. Sci. 30, 545 (2000).
[2] I. Utke, P. Hoffmann, and J. Melngailis, J. Vac. Sci. Technol. B 11, 2386 (2008).
[3] I. S. Beloborodov, A. V. Lopatin, V. M. Vinokur, and K. B. Efetov, Rev. Mod. Phys. 79, 469 (2007).
[4] Y. C. Sun, S. S. Yeh, and J. J. Lin JJ, Phys. Rev. B 82, 054203 (2010).
[5] A. Gerber, A. Milner, G. Deutscher, M. Karpovsky, and A. Gladkikh, Phys. Rev. Lett. 78, 4277 (1997).
[6] K. B. Efetov, and A. Tschersich, Europhys. Lett. 59, 114 (2002).
[7] I. S. Beloborodov, K. B. Efetov, A. V. Lopatin, and V. M. Vinokur, Phys. Rev. Lett. 91, 246801 (2003).
[8] W. F. Van Dorp, and C. W. Hagen, J. Appl. Phys. 104 081301 (2008); S. J. Randolph, J. D. Fowlkes, and P. D. Rack, Crit. Rev. Solid State Mater. Sci. 31, 55 (2006).
[9] A. Botman, J. J. L. Mulders, and C. W. Hagen, Nanotechnology 20, 372001 (2009).
[10] H. W. P. Koops, A. Kaya, and M. Weber, J. Vac. Sci. Technol. B 13, 2400 (1995).
[11] Rotkina L, Lin J F and, J. P. Bird, Appl. Phys. Lett. 83, 4426, (2003)
[12] B. W. Smith and D. E. Luzzi, J. Appl. Phys. 90, 3509 (2001).
[13] D. Ugarte, Nature, 359, 707 (1992).
[14] A. Chuvilin, U. Kaiser, E. Bichoutskaia, N. A. Besley, and A. N. Khlobystov, Nat. Chem. 2, 450 (2010).
[15] C. Miko, M. Milas, J. W. Seo, E. Couteau, Barisic N, Gaal R, and Forro L, Appl. Phys. Lett. 83, 4622 (2003).
[16] A. Krasheninnikov and K. Nordlund, J. Appl. Phys. 107, 071301 (2010).
[17] F. Porrati, R. Sachser, M. Strauss, I. Andrusenko, T. Gorelik, U. Kolb, L. Bayarjargal, B. Winkler, and M. Huth, Nanotechnology 21, 375302 (2010).
[18] A. L. Efros and B. I. Shklovskii, J. Phys. C 8, L49 (1975).
[19] M. Huth, D. Klingenberger, Ch. Grimm, F. Porrati, and R. Sachser, New J. Phys. 11, 033032 (2009).
[20] Y. Tsukatani, N. Yamasaki, K. Murakami, F. Wakaya, and M. Takai, Jpn. J. Appl. Phys., Part 1 44, 5683 (2005).
[21] J. M. De Teresa, R. Cordoba, A. Fernandez-Pacheco, O. Montero, P. Strichovanec, and M. R. Ibarra, J. Nanomater. 2009, 936863 (2009).
[22] R. Sachser, F. Porrati, and M. Huth, Phys. Rev. B 80, 195416 (2009).
[23] F. Porrati, R. Sachser, and M. Huth, Nanotechnology 20, 195301 (2009).

[24] A. C. Ferrari and J. Robertson, Phys. Rev. B 61, 14095 (2000).
[25] N. W. Khun, E. Liu, G. C. Yang, W. G. Ma, and S. P. Jiang, J. Appl. Phys. 106, 013506 (2009).
[26] P. C. Hoyle, M. Ogasawara, J. R. A. Cleaver and H. Ahmed, Appl. Phys. Lett. 62, 3043 (1993).
[27] A. Botman, J. J. L. Mulders, R. Weemaes, and S Mentink, Nanotechnology 17, 3779 (2006).
[28] I. Utke, P. Hoffmann, B. Dwir, K. Leifer, E. Kapon, and P. Doppelt, J. Vac. Sci. Technol. B 18, 3168 (2000).
[29] A. Botman, C. W. Hagen, J. Li, B. L. Thiel, K. A. Dunn, J. J. L. Mulders, S. Randolph, and M. Toth, J. Vac. Sci. Technol. A 27, 2759 (2009).
[30] J. Li, M. Toth, V. Tileli, K. A. Dunn, C.A. Lobo and B. L. Thiel, Appl. Phys. Lett. 93, 023130 (2008).
[31] M. Takeguchi, M. Shimojo, and K. Furuya, Appl. Phys. A 93, 439 (2008).
[32] G. Xie, M. Song, K. Mitsuishi, and Furuya K, Physica E 29, 564 (2005)
[33] J. Li, M. Toth, K. A. Dunn, and B. L. Thiel, J. Appl. Phys. 107, 103540 (2010).
[34] S. Liang, A. Yajima, S. Abe, Y. Mera, and K. Maeda, Surf. Sci. 593, 161 (2005).
[35] A. C. Ferrari, A. Libassi, B. K. Tanner, V. Stolojan, J. Yuan, L. M. Brown, S. E. Rodil, B. Kleinsorge, and J. Robertson, Phys. Rev. B 62, 11089 (2000).
[36] S. Frabboni, G.C. Gazzadi, L. Felisari, and A. Spessot, Appl. Phys. Lett. 88, 213116 (2006).
[37] S. J. Randolph, J. D. Fowlkes, and P. D. Rack, J. Appl. Phys. 97, 124312 (2005).
[38] F. Banhart, Rep. Prog. Phys. 62, 1181 (1999).
[39] M. Huth, J. Appl. Phys. 107, 113709 (2010).
[40] Ch. Schwalb, C. Grimm, M. Baranowski, R. Sachser, F. Porrati, H. Reith, P. Das, J. Müller, F. Völklein, A. Kaya, and M. Huth, Sensor 10, 9847 (2010).

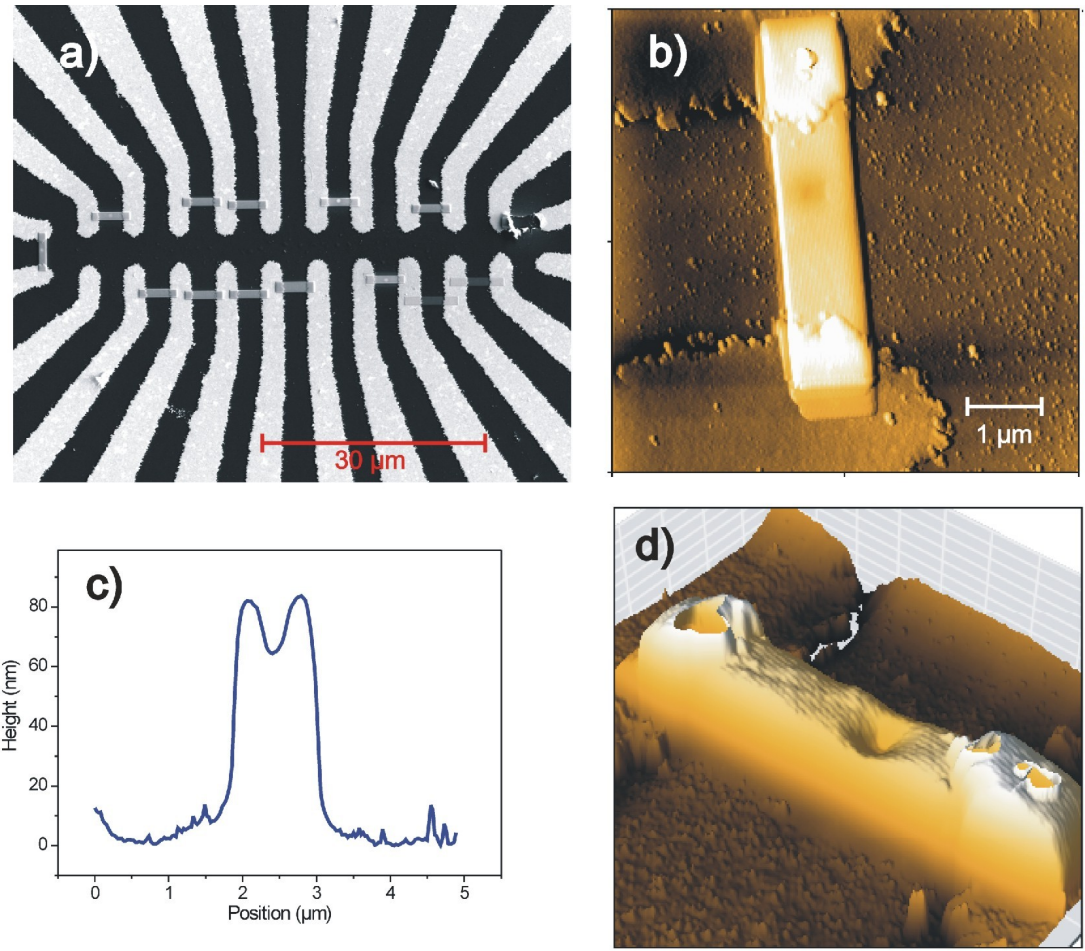

Figure 1: a) SEM picture of the deposits prepared for two-probe electrical measurements. b) AFM image of one deposit. The area is 1x5 µm2. The distance between the electrodes is around 3 µm. c) AFM line scan across the deposit of picture "b". The deposit was locally irradiated to perform an EDX measurement. The effect of the irradiation causes a decrease of the thickness from around 80 nm to 60 nm. d) The depression of the high appear also in the 3D AFM picture.

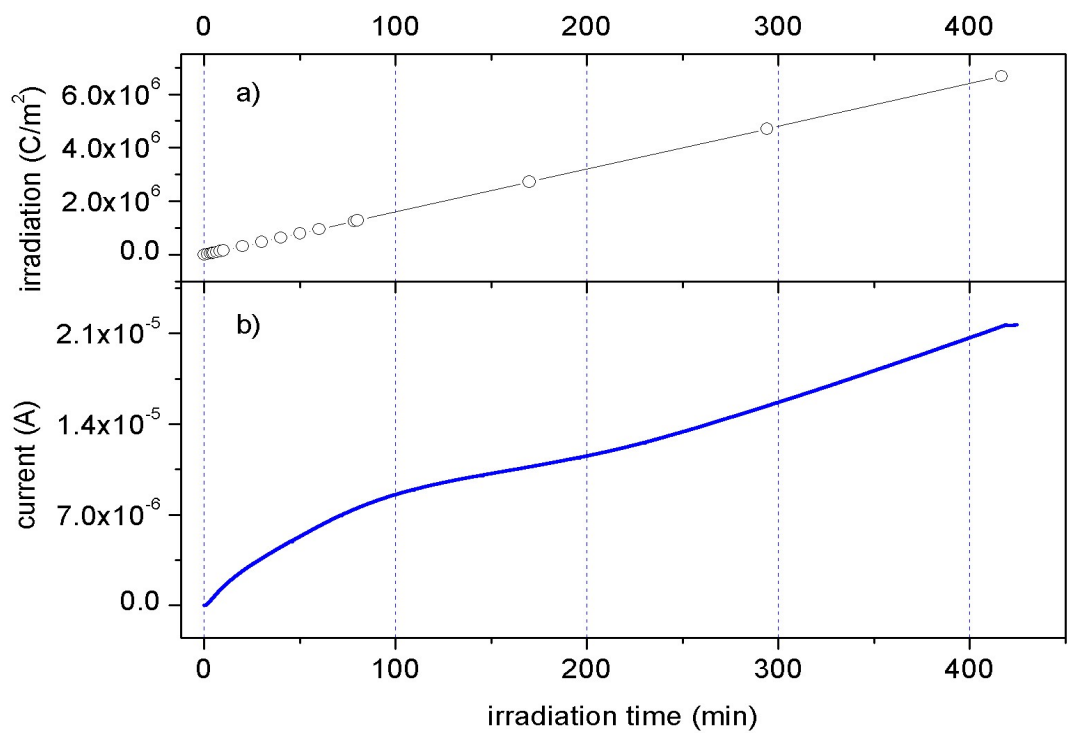

Figure 2: Upper panel: Electron irradiation dose used for the postgrowth irradiation experiment. Each point corresponds to the irradiation dose used for one sample. The irradiation time varies from 100 seconds to 7 hours. Lower panel: Current vs. irradiation time measured for the sample irradiated 7 hours. The applied voltage was 10 mV.

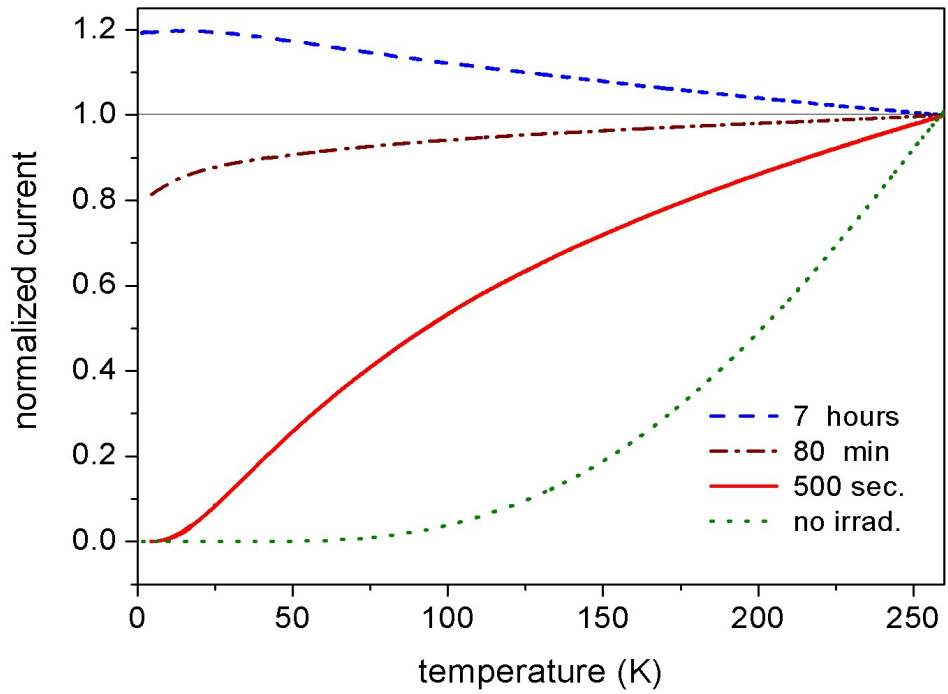

Figure 3: Temperature-dependent conductivity, normalized at 260 K. Green curve (nonirradiated sample): the sample behaves like an insulator. Red curve (500 s irradiation time): the sample is close to the MIT. Brawn curve (80 min irradiated): "metallic" side of the MIT. Blue curve (7 hour irradiation time): the behavior is the one of a metal.

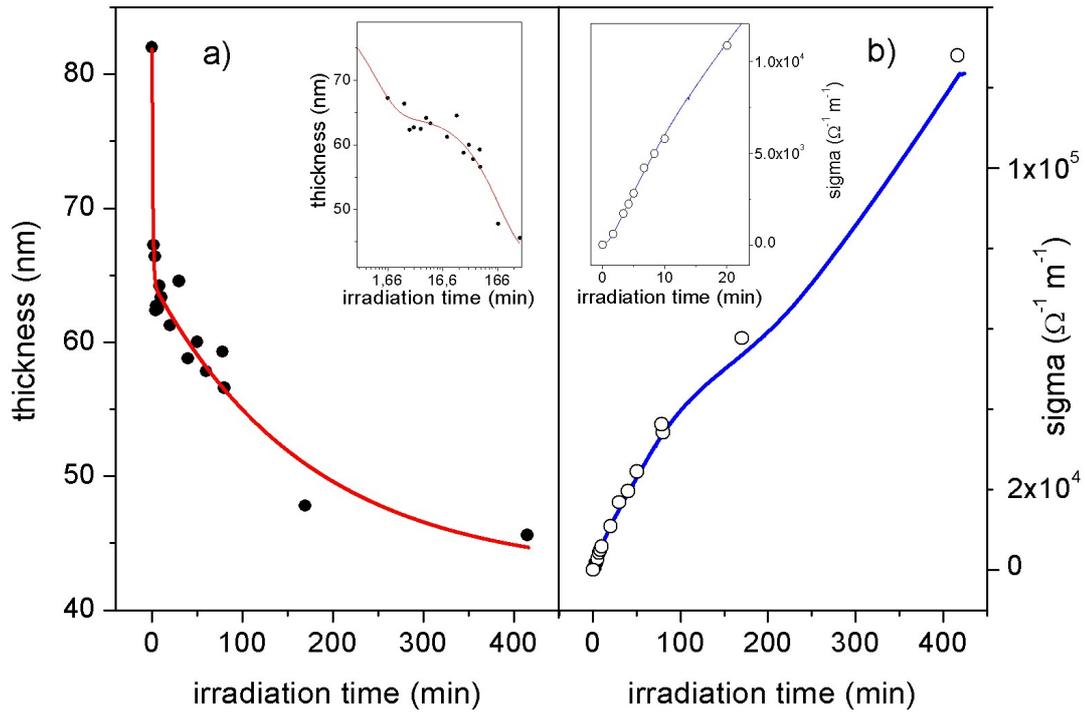

Figure 4: Left panel: Thickness vs. irradiation time. Each point correspond to one sample. The fit is given by the sum of two exponentials. Right panel: Conductivity vs. irradiation time. The blue curve refers to the sample irradiated for 7 hours. The conductivity is given by using the thickness of the sample as deduced by the fit in the left panel. The dots refers to each single sample.

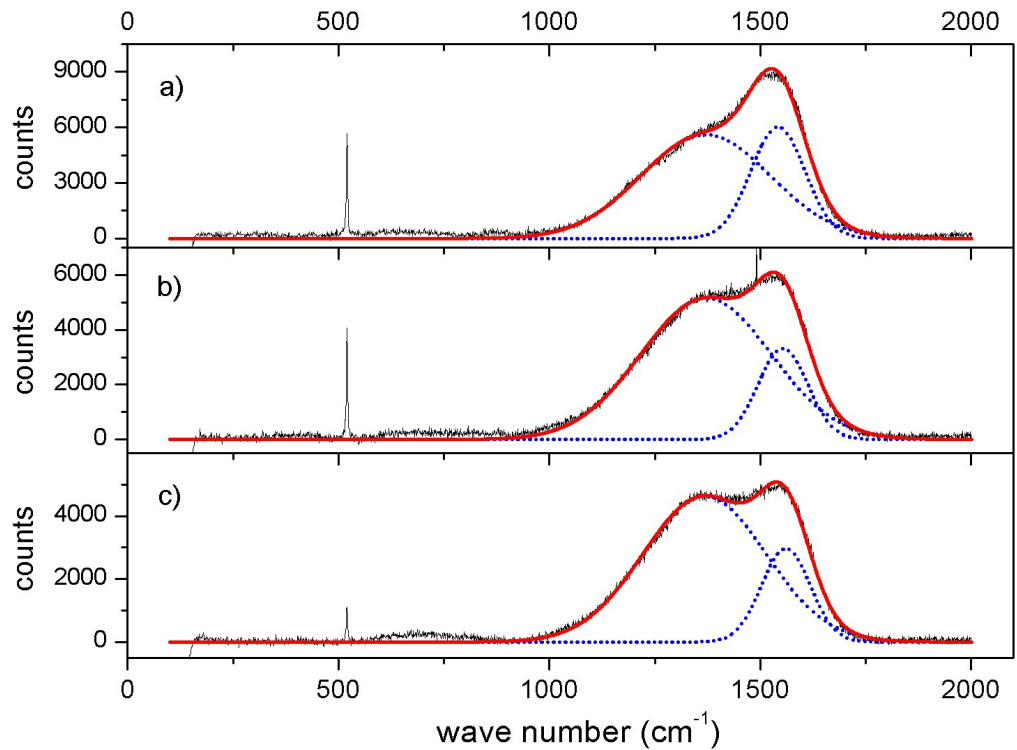

Figure 5: Micro-Raman spectra for: (a) non-irradiated sample, irradiated samples with 1.42x105 C/m2 dose (b) and 2.84x105 C/m2 dose (c). Each spectrum has three peaks at 520 cm−1 (Si substrate background), around 1375 cm−1 ("D" peak) and 1550 cm−1 ("G" peak), respectively. The red curve is a fit to the data by using the gaussian curves (blue) around the "D" and "G" peaks.

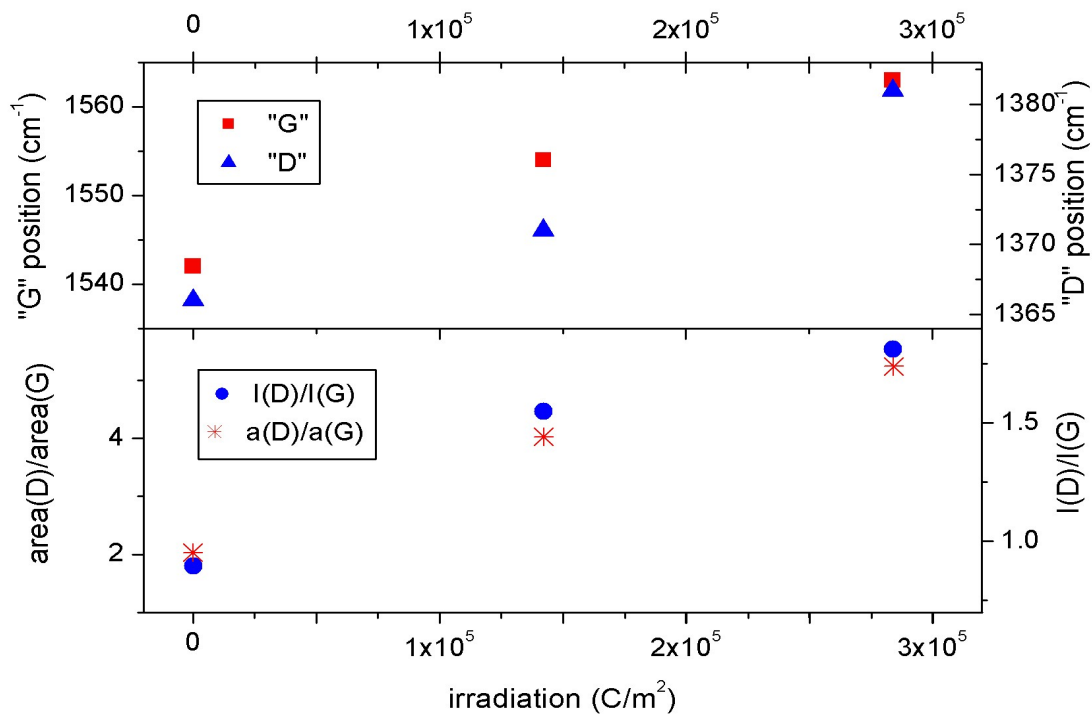

Figure 6: Upper panel: Position of the "D" and "G" peaks vs. irradiation time; the values refer the spectra of Fig 5. Lower panel: Ratio of the intensity (area) of peaks "D" and "G" vs. irradiation time. The peak positions and the intensity ratios give information about the quality of the matrix. In particular, the matrix transform from amorphous carbon to graphite-like with increasing irradiation time.

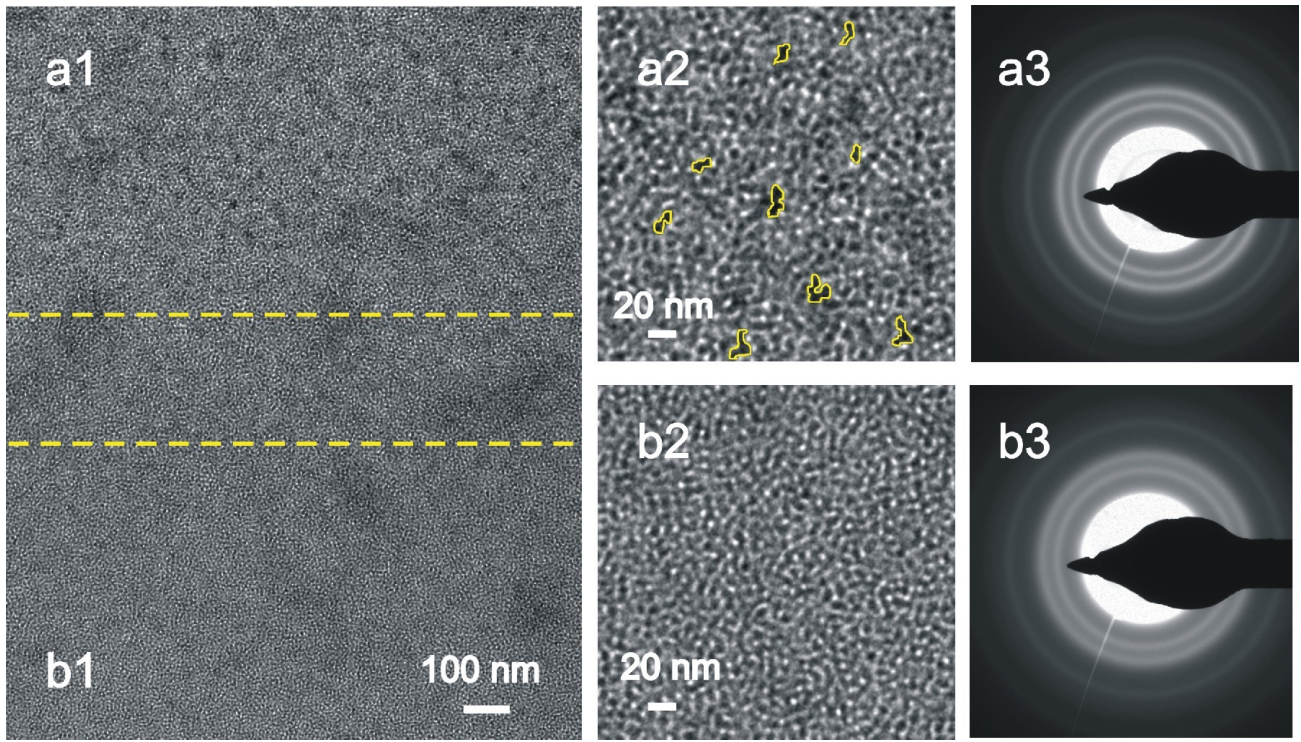

Figure 7: TEM images and corresponding diffraction patterns of the longest irradiated sample. On the left: a1) irradiated area; b1) non-irradiated area. The transition region is marked by the dotted line. Center: selected zoom of the irradiated area (a2) and nonirradiated area (b2). In a2) the coalescence of some crystallites is marked by the yellow contour. On the right: diffraction patterns a3) irradiated area; b3) non-irradiated area.